\documentclass[preprint,showpacs,preprintnumbers]{revtex4}
\usepackage{amsmath}

\begin{document}

\title{Exactly and Quasi-Exactly Solvable two-mode Bosonic Hamiltonians}
\author{Ramazan Ko\c{c}}
\author{Hayriye T\"{u}t\"{u}nc\"{u}ler}
\author{Eser Ol\u{g}ar}
\email{koc@gantep.edu.tr}
\email{tutunculer@gantep.edu.tr}
\email{olgar@gantep.edu.tr}
\affiliation{Department of Physics, Faculty of Engineering\\
University of Gaziantep, 27310 Gaziantep Turkey}
\date{\today}

\begin{abstract}
We develop a method to determine the eigenvalues and eigenfunctions of
two-boson Hamiltonians include a wide class of quantum optical models. The
quantum Hamiltonians have been transformed in the form of the one variable
differential equation and the conditions for its solvability have been
discussed. Applicability of the method is demonstrated on some simple
physical systems.
\end{abstract}

\maketitle











\section{\label{sec:level1}Introduction}

For over ten years there has been a great deal of interest in quantum
optical models which reveal new physical phenomena described by the
Hamiltonians expressed as nonlinear functions of Lie algebra generators or
boson operators\cite{elber,kara1,kara2,delgado,klim}. Such systems have
often been analyzed by using numerical methods, because the implementation
of the Lie algebraic techniques to solve those problems is not very
efficient and most of the other analytical techniques do not yield simple
analytical expressions. They require tedious calculations\cite%
{kara3,band,jurco}.

However, recently a new algebraic approach, essentially improving both
analytical and numerical solution of the problems, has been suggested and
developed for some nonlinear quantum optical systems\cite%
{klim2,kara4,gab,alv}. Most of such developments are mainly based on linear
Lie algebras, but it is evident that there is no physical reason for
symmetries to be only linear. Nonlinear Lie algebra techniques and their
relations to the nonlinear quantum optical systems have been discussed\cite%
{beck,tjin,abd,sunil}. In both cases finite part of spectrum of the
corresponding Hamiltonian can be exactly obtained in closed forms and these
systems are known as quasi-exactly-solvable(QES) termed by Turbiner and
Ushveridze\cite{turb1}. Recently it has been proven that the single boson
Hamiltonians also lead to a QES under some certain constraints\cite%
{dolya1,dolya2}.

The aim of this paper is to determine quasi-exact-solvability of the
two-boson systems and discuss their possible applications in physics. As a
particular case our model includes the solutions of the Karassiov-Klimov
Hamiltonian and the Hamiltonian of the systems of photons and bosons
expressed in a single mode form. These Hamiltonians are not only
mathematically interesting but they have also potential interest in physics%
\cite{per,baj,qu}.

The paper is organized as follows: In section 2, general form of the
Hamiltonian has been constructed and its solution by using the invariance of
the number operator have been discussed. This section includes some physical
examples. In section 3 we present a transformation procedure in order to
solve a wide range of the Hamiltonian. In section 4 the Hamiltonian is
transformed in the form of the one variable differential equation and the
conditions to obtain its eigenvalues and eigenfunctions are discussed.
Finally, \ in section 5 we comment on the validity of our method and suggest
the possible extensions of the problem.

\section{Transformation of the two-mode bosonic systems in the form of
single-variable differential equation}

Two mode bosonic Hamiltonians play an important role in nonlinear quantum
optical systems. The Hamiltonians of such systems can be generalized as
follows:

\begin{equation}
H=\sum_{m_{i}}\alpha
_{m_{1,}m_{2},m_{3},m_{4}}(a_{1}^{+})^{m_{1}}(a_{1})^{m_{2}}(a_{2}^{+})^{m_{3}}(a_{2})^{m_{4}}
\label{1}
\end{equation}%
where $\alpha _{i}$ is a constant and $m_{i}$ determines order of the
interaction. The boson creation, $a_{1}$, $a_{2}$, and annihilation $%
a_{1}^{+}$, $a_{2}^{+}$ operators obey the usual commutation relations

\begin{equation}
\left[ a_{1},a_{2}\right] =\left[ a_{1},a_{2}^{+}\right] =\left[
a_{2},a_{1}^{+}\right] =\left[ a_{1}^{+},a_{2}^{+}\right] =0,\quad \left[
a_{1},a_{1}^{+}\right] =\left[ a_{2},a_{2}^{+}\right] =1.  \label{2}
\end{equation}%
We assume that the conserved quantity of the physical system described by
the Hamiltonian $H,$ correspond to the operator:%
\begin{equation}
K=sa_{1}^{+}a_{1}+pa_{2}^{+}a_{2}.  \label{3}
\end{equation}%
Clearly the Hamiltonian conserves the number of particles in the system,
when $H$ commutes with the $K$. Meanwhile, we have mention here, in general,
conserved quantity of (\ref{3}) type appears more general contexts than the
number operator. Indeed a convenient quantum mechanical treatment of the $%
m^{th}$ order interaction of the two-boson model exploits the $SU(2)$
algebra structure of the problem, by expressing the Hamiltonian as a
polynomial of the angular momentum operators whose degree depends on the
degree of interaction. In this case the conserved quantity is the total
angular momentum, $J=j(j+1)$, which can be related to the number operator $%
j=N/2$, where $N=a_{1}^{+}a_{1}+a_{2}^{+}a_{2}$. Therefore the conserved
quantity $J$ crucially depends on the conservation of the number of
particles. The classical motion of the particle takes place in the space of
angular momentum on the sphere of radius $j(j+1)$. To have a physical
insight in to $K$, from other perspective, when the motion of the particle
takes the place on an ellipsoid the conserved quantity $K$ takes the form (%
\ref{3}). The operator $K$ and bosonic operators satisfy the commutation
relations%
\begin{equation}
\lbrack K,a_{1}^{+}]=sa_{1}^{+},\quad \lbrack K,a_{1}]=-sa_{1},\quad \lbrack
K,a_{2}^{+}]=pa_{2}^{+},\quad \lbrack K,a_{2}]=-pa_{2}.  \label{4}
\end{equation}%
If $K$ and $H$ commute, then they have the same eigenfunction. Therefore it
is worth to seek the conditions for the commutation of $H$ and $K$. This can
easily be done by using the commutation relations (\ref{2}) and (\ref{4})
and we obtain the following relation%
\begin{equation}
\lbrack K,H]=\sum_{m_{i}}[(s(m_{1}-m_{2})+p(m_{3}-m_{4})]\alpha
_{m_{1}m_{2}m_{3}m_{4}}(a_{1}^{+})^{m_{1}}(a_{1})^{m_{2}}(a_{2}^{+})^{m_{3}}(a_{2})^{m_{4}}.
\label{7}
\end{equation}%
It is obvious that the constant of motion $K$ and $H$ commute when the
following set of equation is satisfied%
\begin{equation}
s(m_{1}-m_{2})+p(m_{3}-m_{4})=0.  \label{8}
\end{equation}%
The action of the operator $K$ on the state $\left| n_{1},n_{2}\right\rangle 
$ is given by%
\begin{equation}
K\left| n_{1},n_{2}\right\rangle =(sn_{1}+pn_{2})\left|
n_{1},n_{2}\right\rangle .  \label{5}
\end{equation}%
In this section we seek the solution of the eigenvalue equation (\ref{5}) in
the Bargmann-Fock space. The usual realization of the bosonic operators in
Hilbert space is given by 
\begin{subequations}
\begin{eqnarray}
a_{1}^{+} &=&\frac{1}{\sqrt{2}}\left( -\frac{\partial }{\partial x_{1}}%
+x_{1}\right) ;\quad a_{1}=\frac{1}{\sqrt{2}}\left( \frac{\partial }{%
\partial x_{1}}+x_{1}\right)   \label{q1} \\
a_{2}^{+} &=&\frac{1}{\sqrt{2}}\left( -\frac{\partial }{\partial x_{2}}%
+x_{2}\right) ;\quad a_{2}=\frac{1}{\sqrt{2}}\left( \frac{\partial }{%
\partial x_{2}}+x_{2}\right)   \label{q2}
\end{eqnarray}%
The operators can be transformed to the Bargmann-Fock space by introducing
the operator 
\end{subequations}
\begin{equation}
\Gamma =\exp \left[ \frac{\pi }{8}\left(
a_{1}^{2}+a_{1}^{+2}+a_{2}^{2}+a_{2}^{+2}\right) \right]   \label{q3}
\end{equation}%
and the similarity transformation%
\begin{eqnarray}
b_{1} &=&\Gamma a_{1}\Gamma ^{-1}=\frac{1}{\sqrt{2}}\left(
a_{1}^{+}-a_{1}\right) ,\quad b_{1}^{+}=\Gamma a_{1}^{+}\Gamma ^{-1}=\frac{1%
}{\sqrt{2}}\left( a_{1}^{+}+a_{1}\right)   \notag \\
b_{2} &=&\Gamma a_{2}\Gamma ^{-1}=\frac{1}{\sqrt{2}}\left(
a_{2}^{+}-a_{2}\right) ,\quad b_{2}^{+}=\Gamma a_{2}^{+}\Gamma ^{-1}=\frac{1%
}{\sqrt{2}}\left( a_{2}^{+}+a_{2}\right) .  \label{q4}
\end{eqnarray}%
The Bargmann-Fock space differential realizations of the operators take the
form%
\begin{equation}
b_{1}=\frac{\partial }{\partial x_{1}};\quad b_{1}^{+}=x_{1};\quad b_{1}=%
\frac{\partial }{\partial x_{2}};\quad b_{2}^{+}=x_{2}  \label{qb}
\end{equation}%
Thus, in the Bargmann-Fock space the eigenvalue problem (\ref{5}) leads to
the following solution:%
\begin{equation}
\psi (x_{1},x_{2})=x_{1}^{k}\phi \left( x_{2}x_{1}^{-\frac{s}{p}}\right) .
\label{6}
\end{equation}%
where $k$ is given by%
\begin{equation}
k=n_{1}+\frac{p}{s}n_{2}.  \label{6a}
\end{equation}%
The eigenfunction of the Hamiltonian can be obtained from the relation%
\begin{equation}
\left| n_{1},n_{2}\right\rangle =\Gamma ^{-1}\psi (x_{1},x_{2}).  \label{q5}
\end{equation}%
In the following section the application of our procedure will be discussed
explicitly.

\subsection{Hamiltonian of the second Harmonic generation and generalization
of the solvability conditions}

The Hamiltonian (\ref{1}) includes various physical Hamiltonians, second and
third harmonic generation effective Hamiltonians. This gives us an
opportunity to test our approach because those Hamiltonians have been
studied in literature. Consider the following Hamiltonian:%
\begin{equation}
H=\omega _{1}a_{1}^{+}a_{1}+\omega _{2}a_{2}^{+}a_{2}+\kappa
(a_{1}^{+})^{2}a_{2}+\overline{\kappa }a_{2}^{+}(a_{1})^{2}  \label{11}
\end{equation}
which can be related to the Hamiltonian (\ref{1}), when the parameters%
\begin{equation}
\alpha _{m_{1}m_{2},m_{3},m_{4}}=0  \label{9}
\end{equation}%
except that%
\begin{equation}
\alpha _{1,1,0,0}=\omega _{1},\alpha _{0,0,1,1}=\omega _{2},\alpha
_{2,0,0,1}=\kappa ,\alpha _{0,2,1,0}=\overline{\kappa }  \label{10}
\end{equation}%
The condition (\ref{8}) is satisfied when $p=2s$. The constants $\omega _{1}$
and $\omega _{2}$ are the angular frequencies of two independent harmonic
oscillators characterized by annihilation and creation operators, $\kappa $
and $\overline{\kappa }$ are the coupling coefficients that determine the
strength of the interaction of the oscillators. In the following we use the
Bargmann-Fock realization, where creation and annihilation operators $%
a_{i}^{+}$ and $a_{i}$ are replaced by $b_{i}^{+}$ and $b_{i}$,
respectively. The eigenfunction (\ref{6}) is of the form

\begin{equation}
\psi (x_{1},x_{2})=(x_{1})^{k}\phi (z)  \label{13}
\end{equation}%
where $z=x_{2}x_{1}^{-1/2}$. The solution of this system describes a quantum
mechanical state of $H$ provided that $\phi (z)$ belongs to the
Bargmann-Fock space. The eigenvalue equation of the second harmonic
generation Hamiltonian can be written as%
\begin{equation}
H\psi (x_{1},x_{2})=E\psi (x_{1},x_{2})  \label{15}
\end{equation}%
Insertion of (\ref{13}) into (\ref{15}) yields the following differential
equation

\begin{equation}
\left[ 4\overline{\kappa }z^{3}\frac{d^{2}}{dz^{2}}+(\kappa +z(\omega
_{2}-2\omega _{1}+2\overline{\kappa }z(3-2k))\frac{d}{dz}+\omega
_{2}+k\omega _{1}+\overline{\kappa }zk(k-1)-E\right] \phi (z)=0.  \label{16}
\end{equation}%
According to Turbiner\cite{turb1} the Hamiltonian (\ref{16}) is
quasi-exactly solvable. The eigenvalue equation (\ref{16}) can be obtained
as follows. Let us assume that the function $\phi (z)$ is a polynomial in $z$
with the coefficients being functions of energy:

\begin{equation}
\phi (z)=\sum\limits_{m=0}^{\infty }P_{m}(E)z^{m}.  \label{17}
\end{equation}%
When inserted $\phi (z)$ in (\ref{16}) then we obtain the following
three-term recurrence relation:

\begin{eqnarray}
\overline{\kappa }(k-2m)(k-2m-1)P_{m+1}(E)+ &&  \label{18} \\
(\omega _{2}+\kappa \omega _{1}+m(\omega _{2}-2\omega
_{1})-E)P_{m}(E)+\kappa mP_{m-1}(E) &=&0.  \notag
\end{eqnarray}%
The function $P_{m}(E)$ terminates when $m>k/2$, and therefore the roots of
the $P_{m}(E)$ belong to the spectrum of the Hamiltonian. For the
corresponding problem, Beckers, Brihaye and Debergh\cite{beck} have obtained
three different Hamiltonians in the framework of nonlinear algebra. Two of
the Hamiltonians are PT- symmetric and they related to the sextic harmonic
oscillator. The third Hamiltonian is exactly the same one in (\ref{16})
which was obtained in a different method. The Hamiltonian (\ref{16}) can
also be transformed in the form of the Schr\"{o}dinger equation by changing
variable $z=-1/\overline{\kappa }y^{2}$ and defining the wavefunction%
\begin{equation}
\phi (z)=e^{-\int W(y)dy}\psi (y)  \label{19}
\end{equation}%
where%
\begin{equation}
W(y)=\frac{k}{y}+\frac{(\omega _{2}-2\omega _{1})y}{4}-\frac{\kappa 
\overline{\kappa }y^{3}}{4}  \label{20}
\end{equation}%
Substituting (\ref{19}) in (\ref{16}) we obtain the following equation:

\begin{eqnarray}
H &=&-\frac{1}{2}\frac{d^{2}}{dy^{2}}+\frac{1}{4}\left( (2k+5)\omega
_{2}-2\omega _{1})\right) +\frac{1}{16}\left( (\omega _{2}-2\omega
_{1})^{2}-4\kappa \overline{\kappa }(2k+3)\right) y^{2}-  \notag \\
&&\frac{1}{8}\kappa \overline{\kappa }(\omega _{2}-2\omega _{1})y^{4}+\frac{1%
}{16}\kappa ^{2}\overline{\kappa }^{2}y^{6}.  \label{21}
\end{eqnarray}%
The last equation is the Hamiltonian of the sextic Harmonic oscillator.
Consequently the procedure given here can be applied to obtain eigenfunction
and eigenvalues of the various physical Hamiltonians. The validity of the
procedure depends on the choice of the $\alpha _{m_{1}m_{2},m_{3},m_{4}}.$%
One can easily be obtain various physical Hamiltonians by appropriate choice
of $\alpha _{m_{1}m_{2},m_{3},m_{4}}$ and by considering the conditions
given in (\ref{8}). For instance one can easily obtain $n^{th}$ harmonic
generation Hamiltonian by setting:%
\begin{equation}
\alpha _{1,1,0,0}=\omega _{1},\alpha _{0,0,1,1}=\omega _{2},\alpha
_{n,0,0,1}=\kappa ,\alpha _{0,n,1,0}=\overline{\kappa }  \label{22}
\end{equation}%
and the remaining values of the $\alpha _{m_{1}m_{2},m_{3},m_{4}}=0$. Then
the Hamiltonian (\ref{1}) takes the form%
\begin{equation}
H=\omega _{1}a_{1}^{+}a_{1}+\omega _{2}a_{2}^{+}a_{2}+\kappa
(a_{1}^{+})^{n}a_{2}+\overline{\kappa }a_{2}^{+}(a_{1})^{n}.  \label{23}
\end{equation}%
In the following section we continue to explore the solution of the
Hamiltonian (\ref{1}), in the Bargmann-Fock space by developing a similarity
transformation procedure.

\section{Transformation of the bosonic operators}

In this section we present two transformation procedure to obtain the
conditions of the solvability of the Hamiltonian (\ref{1}). The following
transformations allow us to study a wide range of physical systems. Let us
introduce the following similarity transformation induced by the operator

\begin{equation}
S=(a_{2}^{+})^{\eta a_{1}^{+}a_{1}}  \label{24}
\end{equation}%
where $\eta $ is a constant and it can be determined by considering the
transformation of the number operator (\ref{3}). The operator $S$ acts on
the state $\left| n_{1},n_{2}\right\rangle $ as follows,

\begin{equation}
S\left| n_{1},n_{2}\right\rangle =(a_{2}^{+})^{\eta n_{1}}\left|
n_{1},n_{2}\right\rangle =\sqrt{\frac{n_{2}!}{(n_{2}+\eta n_{1})!}}\left|
n_{1},n_{2}+\eta n_{1}\right\rangle .  \label{25}
\end{equation}%
Since $a_{1}$ and $a_{2}$ commute, the transformation of $a_{1}$ and $%
a_{1}^{+}$ under $S$ can be obtained by writing $a_{2}^{+}=e^{b}$, with $%
[a_{1},b]=[a_{1}^{+},b]=0$,

\begin{eqnarray}
Sa_{1}^{+}S^{-1} &=&e^{\eta ba_{1}^{+}a_{1}}a_{1}^{+}e^{-\eta
ba_{1}^{+}a_{1}}=a_{1}^{+}(a_{2}^{+})^{\eta }  \notag \\
Sa_{1}S^{-1} &=&e^{\eta ba_{1}^{+}a_{1}}a_{1}e^{-\eta
ba_{1}^{+}a_{1}}=a_{1}(a_{2}^{+})^{-\eta }  \label{26}
\end{eqnarray}%
and transformation of $a_{2}$ and $a_{2}^{+}$ as follows

\begin{eqnarray}
Sa_{2}^{+}S^{-1} &=&(a_{2}^{+})^{\eta
a_{1}^{+}a_{1}}a_{2}^{+}(a_{2}^{+})^{-\eta a_{1}^{+}a_{1}}=a_{2}^{+}  \notag
\\
Sa_{2}S^{-1} &=&(a_{2}^{+})^{\eta a_{1}^{+}a_{1}}a_{2}(a_{2}^{+})^{-\eta
a_{1}^{+}a_{1}}=a_{2}-\eta a_{1}^{+}a_{1}(a_{2}^{+})^{-1}.  \label{27}
\end{eqnarray}%
In a similar manner, we prepare the other similarity transformation which is
useful to study QES of the Hamiltonian (\ref{1}), by introducing the
following operator:

\begin{equation}
T=a_{2}^{\alpha a_{1}^{+}a_{1}}  \label{28}
\end{equation}%
The operator $T$ acts on the two-boson state as

\begin{equation}
T\left| n_{1},n_{2}\right\rangle =a_{2}^{\alpha n_{1}}\left|
n_{1},n_{2}\right\rangle =\sqrt{\frac{n_{2}!}{(n_{2}-\alpha n_{1})!}}\left|
n_{1},n_{2}-\alpha n_{1}\right\rangle .  \label{29}
\end{equation}%
Since $a_{1}$ and $a_{2}$ commute, the transformation of $a_{1}$ and $%
a_{1}^{+}$ under $S$ can be obtained by letting $a_{2}=e^{c}$ with $%
[a_{1},c]=[a_{1}^{+},c]=0,$

\begin{eqnarray}
Ta_{1}^{+}T^{-1} &=&e^{\alpha ca_{1}^{+}a_{1}}a_{1}^{+}e^{-\alpha
ca_{1}^{+}a_{1}}=a_{1}^{+}(a_{2})^{\alpha }  \notag \\
Ta_{1}T^{-1} &=&e^{\alpha ca_{1}^{+}a_{1}}a_{1}e^{-\alpha
ca_{1}^{+}a_{1}}=a_{1}(a_{2})^{-\alpha }  \label{30}
\end{eqnarray}%
The transformation of $a_{2}$ and $a_{2}^{+}$ is as follows:

\begin{eqnarray}
Ta_{2}^{+}T^{-1} &=&a_{2}^{\alpha a_{1}^{+}a_{1}}a_{2}^{+}a_{2}^{-\alpha
a_{1}^{+}a_{1}}=a_{2}^{+}+\alpha a_{1}^{+}a_{1}a_{2}^{-1}.  \notag \\
Ta_{2}T^{-1} &=&a_{2}^{\alpha a_{1}^{+}a_{1}}a_{2}a_{2}^{-\alpha
a_{1}^{+}a_{1}}=a_{2}  \label{31}
\end{eqnarray}%
These two transformations lead to the conditions of the
quasi-exact-solvability of the Hamiltonian (\ref{1}).

\section{Solvability of the Hamiltonian}

In this section we discuss the solvability of the Hamiltonian(\ref{1}). The
conserved quantity $K$ of the physical system describes the states of the
corresponding Hamiltonian. The transformation of $K$ under the operator $S$
is given by%
\begin{equation}
K^{\prime }=SKS^{-1}=(s-p\eta )a_{1}^{+}a_{1}+pa_{2}^{+}a_{2}  \label{32}
\end{equation}%
The Hamiltonian (\ref{1}) is characterized by the total number of $a_{1}$
and $a_{2}$ bosons when the conserved quantity of the physical system $K$
commutes with the whole Hamiltonian. In the transformed case it is only the
number of $a_{2}$ bosons that characterize the system under the condition $%
\eta =s/p$. According to (\ref{32}) the representation is characterized by a
fixed number $a_{2}^{+}a_{2}=k$. Therefore in the transformed form the
Hamiltonian can be expressed as one boson operator $a_{1}$, when the
condition (\ref{8}) is taken into consideration. The transformed form of the
Hamiltonian (\ref{1}) can be written as:%
\begin{equation}
\widetilde{H}=SHS^{-1}=\sum_{m_{i}}\alpha
_{m_{1,}m_{2,}m_{4}}(a_{1}^{+})^{m_{1}}(a_{1})^{m_{2}}(k-\frac{s}{p}%
a_{1}^{+}a_{1})^{m_{4}}  \label{33}
\end{equation}%
The difference between (\ref{1}) and (\ref{33}) is that while in the first
the total number of $a_{1}$ and $a_{2}$ bosons characterize the the system,
in the later it is only the number of $a_{2}$ bosons that characterize the
system. Therefore the representation is characterized by a fixed number $k$
and in (\ref{33}), the Hamiltonian is expressed in terms of one boson
operator $a_{1}$. The transformed Hamiltonian $\widetilde{H}$, in the
Bargmann-Fock space, which plays an important role in the quasi-exact
solution of the equation (\ref{1}). It can be transformed in the form of the
one dimensional differential equations in the Bargmann-Fock space when the
boson operators are realized as%
\begin{equation}
a_{1}=\frac{d}{dx},\quad a_{1}^{+}=x.  \label{34}
\end{equation}%
The basis function of the primed generators of the system is the degree of
polynomial of order $k$,

\begin{equation}
P_{n}(x)=(x^{0},x^{1},\cdots ,x^{k}).  \label{35}
\end{equation}%
Action of (\ref{33}) on the (\ref{35}), in the Bargmann-Fock space can be
written as:%
\begin{equation}
\widetilde{H}P_{n}(x)=\sum P_{n}(E)x^{n}  \label{36}
\end{equation}%
where $E$ is the eigenfunction of the Hamiltonian and the polynomial $P(E)$
can be written as%
\begin{equation}
\sum_{m_{i}}\alpha _{m_{1,}m_{2,}m_{4}}(k-\frac{s}{p}n)^{m_{4}}\frac{n!}{%
(n-m_{2})!}P_{n+m_{1}-m_{2}}(E)-EP_{n}(E)=0  \label{37}
\end{equation}%
The wavefunction is itself the generating function of the energy
polynomials. The eigenvalues are then produced by the roots of such
polynomials. If the $E_{j}$ is a root of the polynomial $P_{j+1}(E)$, the
series (\ref{37}) terminates at $j>k\frac{p}{s}$ and $E_{j}$ belongs to the
spectrum of the corresponding Hamiltonian. The eigenvalues are then obtained
by finding the roots of such polynomials.

The constant of motion $K$ characterize the system can be transformed, by
the operator $T$ is given by%
\begin{equation}
K^{\prime }=TKT^{-1}=(s+p\eta )a_{1}^{+}a_{1}+pa_{2}^{+}a_{2}.  \label{38}
\end{equation}%
The Hamiltonian (\ref{1}) can be characterized, in the transformed case, $%
\eta =-s/p$. Thus according to (\ref{38}) the representation is
characterized by a fixed number $a_{2}^{+}a_{2}=k$. Therefore the
transformed Hamiltonian includes one boson operator $a_{1}$, when the
condition (\ref{8}) is taken into consideration, and it can be written as:%
\begin{equation}
H^{\prime }=THT^{-1}=\sum_{m_{i}}\alpha
_{m_{1,}m_{2,}m_{4}}(a_{1}^{+})^{m_{1}}(a_{1})^{m_{2}}(k+\frac{s(m_{1}-m_{2})%
}{pm_{3}}+\frac{s}{p}a_{1}^{+}a_{1})^{m_{3}}  \label{39}
\end{equation}%
The Hamiltonian can be expressed as one dimensional differential equation in
the Bargmann-Fock space. In order to solve the Hamiltonian $H^{\prime }$ we
can follow the procedure given (\ref{34}) through (\ref{37}). In this case
the polynomial function is terminated when $k$ is constrained to%
\begin{equation}
k>-\frac{s(m_{1}-m_{2})}{pm_{3}}-\frac{s}{p}n  \label{40}
\end{equation}%
Consequently we have obtained two classes of Hamiltonians whose spectrum can
be obtained (quasi)exactly.

\section{Conclusion}

In this paper we have prepared a general method to obtain the solution of
two boson Hamiltonian. By using either solution of number operator or
similarity transformation, we have been able to provide a QES of the various
physical Hamiltonians. Furthermore, it has been given that two boson
Hamiltonian can be reduced to single variable differential equation in the
Bargmann-Fock space. It is also important to mention here that the methods
given here can be used to solve higher order differential equations.

The method given here, can easily be extended to solve the Hamiltonians that
include multi-boson or fermion-boson systems. This extension leads to the
solution of the various physical problems, such as Pauli equation, Rabi
Hamiltonian, Jaynes-Cummings Hamiltonian.

\end{document}